\documentclass[prb,onecolumn,floatfix,showpacs]{revtex4}
\usepackage{graphicx,bm}
\begin{document}
\title{Field sweep rate dependence of magnetic domain patterns:
Numerical simulations for a simple Ising-like model}

\author{Kazue Kudo}
\email{kudo@a-phys.eng.osaka-cu.ac.jp}
\author{Katsuhiro Nakamura}
\affiliation{Department of Applied Physics, Graduate School of
Engineering, Osaka City University, Osaka 558-8585, Japan}

\date{\today}

\begin{abstract}
We study magnetic domain patterns in ferromagnetic thin films by
 numerical simulations for a simple Ising-like model. Magnetic domain
 patterns after quench demonstrate various types of patterns
 depending on the 
 field sweep rate and parameters of the model. 
How the domain patterns are formed is shown with the 
use of the number of domains,
 the domain area, and domain-area distributions, as well as snapshots of
 domain patterns.  
Considering the proper time scale of the system, we propose a
 criterion for the structure of domain patterns. 
\end{abstract}


\maketitle

\section{\label{sec:intro} Introduction}

Various kinds of domain pattern are seen in a large number of physical
and chemical systems and have been investigated experimentally,
numerically, 
and theoretically.\cite{cross,seul} Magnetic domain patterns in uniaxial
ferromagnetic garnet thin films also show many types of structure. 
For example,
a hexagonal or square lattice, bubbles, stripes, and so on are observed
under static or oscillating magnetic field perpendicular to the film
plane. Under zero field, after quench, a labyrinth structure is often
observed. In some cases, the domain pattern under zero field has a
sea-island structure, which 
consists of many small up-spin (down-spin) ``islands'' (domains)
surrounded by the ``sea'' of down (up) spins. Labyrinth and sea-island
structures are metastable patterns and should change to a
parallel-stripe structure under the thermal fluctuation or other
kinds of fluctuation. In fact, it was observed in experiments that a
labyrinth structure changes to a parallel-stripe structure under field
cycles.\cite{miura,mino}

Magnetic domain patterns show different structures depending on the
external field. There are several simulations which demonstrate how
domain patterns change under very slowly changing
field.\cite{jagla04,jagla05,deutsch} 
Some of them also showed hysteresis
curves.\cite{jagla05,deutsch} 
Though a hysteresis curve has no information about the structure
of a domain pattern, it gives information about magnetic ordering.
Magnetic ordering depends not only on the external field but also on the
field changing rate. In fact, there are also some works about the
dependence of hysteresis on the frequency and amplitude of an
oscillating field.\cite{luse,sides,jang}

In this paper, we focus on the domain patterns under zero field and
study them by numerical simulations. Our model is a simple Ising-like
model which reproduces domain patterns observed in experiments for
ferromagnetic garnet thin films. In experiments, domain patterns under
zero field are observed after a strong applied field is switched
off. Therefore, we need to start a simulation from the state under
magnetic field above the saturation field in order to reproduce the
domain pattern under zero field.
Recently, the field sweep-rate dependence of domain patterns was
studied experimentally and numerically in Ref.~\onlinecite{kudo}.
However, the parameters except for the field
sweep rate were fixed to specific values in the simulations.
The parameters are different among samples in experiments. 
Here, we will also study a
parameter dependence of domain patterns. The parameter dependence as
well as the field sweep-rate dependence may give useful information on
rich properties of the pattern formation in a ferromagnetic thin film.

The rest of this paper is organized as follows. In Sec.~\ref{sec:model},
we describe our model and numerical procedure. In Secs.~\ref{sec:fast}
and \ref{sec:slow}, the characteristics of domain patterns are displayed
for fast- and slow-quench cases, respectively. We will show not only snapshots
of domain patterns but also the number of domains, domain areas, and the
domain-area distributions. Such quantities enable us to picture how the
domain patterns appear. 
In Sec.~\ref{sec:cri}, a criterion about domain patterns and the field
sweep rate is discussed. 
Conclusions are given in  Sec.~\ref{sec:con}.

\section{\label{sec:model} Model and numerical procedure}

Our model is a simple two-dimensional Ising-like model whose Hamiltonian
consists of four energy terms: 
uniaxial-anisotropy energy $H_{\rm ani}$, local ferromagnetic interactions
(exchange interactions) $H_J$, long range dipolar interactions
$H_{\rm di}$, and interactions with the external field 
$H_{\rm ex}$.\cite{jagla04,jagla05,kudo,elias,sagui} 
We consider a scalar field $\phi(\bm{r})$, where $\bm{r}=(x,y)$. The
positive and negative values of $\phi(\bm{r})$ correspond to up and down
spins, respectively. The anisotropy energy prefers the values
$\phi(\bm{r})=\pm 1$:
\begin{equation}
 H_{\rm ani}=\alpha \int {\rm d}\bm{r} \left(
-\frac{\phi(\bm{r})^2}{2}+\frac{\phi(\bm{r})^4}{4}
\right).
\label{eq:Ha}
\end{equation}
The exchange and dipolar interactions are described by
\begin{equation}
 H_J=\beta\int {\rm d}\bm{r} \frac{|\nabla\phi(\bm{r})|^2}{2}
\label{eq:Hj}
\end{equation}
and
\begin{equation}
 H_{\rm di}=\gamma\int {\rm d}\bm{r} {\rm d}\bm{r}'
 \phi(\bm{r})\phi(\bm{r}') G(\bm{r},\bm{r}'),
\label{eq:Hdi}
\end{equation}
respectively. Here, $G(\bm{r},\bm{r}')\sim |\bm{r}-\bm{r}'|^{-3}$ at
long distances.
The exchange and dipolar interactions may be interpreted as short-range
attractive and long-range repulsive interactions. Namely, $\phi(\bm{r})$
tends to have the same value as neighbors because of the exchange
interactions, and it also tends to have the opposite sign to the values
in a region at a long distance because of the dipolar interactions.
These two interactions play an important role in creating a domain
pattern with a characteristic length.
The term from the interactions with the external field is given by
\begin{equation}
 H_{\rm ex}=-h(t) \int {\rm d}\bm{r} \phi(\bm{r}),
\label{eq:Hex}
\end{equation}
where $h(t)$ is the time-dependent external field. 
Here, let us introduce a disorder effect. 
In other words, spatial randomness is introduced in the model.
We consider the disorder effect only in the anisotropy term: 
$\alpha$ is replaced by $\alpha\lambda(\bm{r})$, where
\begin{equation}
 \lambda(\bm{r})=1+\mu(\bm{r})/4. 
\label{eq:random}
\end{equation}
Here, $\mu(\bm{r})$ is an uncorrelated random number with a Gaussian
distribution whose average and variance are 0 and $\mu_0^2$, respectively.
However, $\mu(\bm{r})$ should have a cutoff so that $\lambda(\bm{r})$
is always positive. In our simulations, we set $\mu_0=0.3$.
From Eqs.~(\ref{eq:Ha})--(\ref{eq:random}), the dynamical equation
of our model is described by 
\begin{eqnarray}
 \frac{\partial \phi (\bm{r})}{\partial t}&=&
 -L_0\frac{\delta (H_{\rm ani}+H_{J}+H_{\rm di}+{H_{\rm ex}})}
 {\delta \phi (\bm{r})} \nonumber\\
&=& L_0 \biggl\{ \alpha \lambda(\bm{r}) [\phi(\bm{r})-\phi(\bm{r})^3]
+\beta\nabla^2\phi(\bm{r})
-\gamma\int {\rm d}\bm{r}' \phi(\bm{r}') G(\bm{r},\bm{r}')
+h(t)\bigg\}.
\label{eq:A-C}
\end{eqnarray}
Hereafter, we fix $L_0=1$.

It is useful to calculate the time evolutions of Eq.~(\ref{eq:A-C}) in
Fourier space when we perform numerical simulations. The equation is
rewritten as 
\begin{equation}
 \frac{\partial \phi_{\bm k}}{\partial t}
 =\alpha  [(\phi-\phi^3)\lambda ]_{\bm{k}}
-(\beta k^2+\gamma G_{\bm{k}})\phi_{\bm{k}}
+h(t)\delta_{\bm{k},0},
\label{eq:k-eq}
\end{equation}
where $[\cdot]_{\bm{k}}$ denotes the convolution sum and $G_{\bm{k}}$
is the Fourier transform of $G(\bm{r},0)$. 
If $G(\bm{r},0)\equiv 1/|\bm{r}|^3$, then
\begin{equation}
 G_{\bm{k}}=a_0-a_1 k,
\label{eq:Gk}
\end{equation}
where $k=|\bm{k}|$ and
\begin{equation}
 a_0=2\pi\int_d^\infty r{\rm d}r G(r), \quad a_1=2\pi.
\label{eq:a0a1}
\end{equation}
Here, $d$ is the cutoff length, which can be interpreted as the lower
limit of the dipolar interactions. In fact, $G(\bm{r},0)$ cannot be assumed
to be $1/|\bm{r}|^3$ at short distances, and Eq.~(\ref{eq:Gk}) is the
asymptotic form for $k\to 0$. However, the details of
$G(\bm{r},0)$ is not important at short distances because the exchange
interactions (the $k^2$ term) are dominant there. In the simulations
below, we use Eqs.~(\ref{eq:Gk}) and (\ref{eq:a0a1}) as $G_{\bm{k}}$ and
set $d=\pi /2$, which results in $a_0=4$.

Equation~(\ref{eq:k-eq}) is useful also for discussing some
characteristic properties
of domain patterns. Let us consider only linear terms in the equation
\begin{equation}
 \frac{\partial \phi_{\bm{k}}}{\partial t}=\eta_{\bm{k}}\phi_{\bm{k}},
\end{equation}
where $\eta_{\bm{k}}$ is the linear-growth rate for zero field. It means
that 
$\phi_{\bm{k}}$ decays exponentially for negative $\eta_{\bm{k}}$ and
grows exponentially for positive $\eta_{\bm{k}}$, although the nonlinear
term prevents $\phi_{\bm{k}}$'s growing too much. The linear-growth rate
has a quadratic maximum:
\begin{eqnarray}
 \eta_{\bm{k}}&=&-(\beta k^2-\gamma a_1k+\gamma a_0)+\alpha \nonumber\\
&=& -\beta \left( k-\frac{a_1\gamma}{2\beta} \right)^2
+\frac{a_1^2\gamma^2}{4\beta}-\gamma a_0+\alpha .
\label{eq:eta}
\end{eqnarray}
This suggests that the characteristic length of the domain patterns
should be $2\pi/k_0$, where $k_0=a_1\gamma/2\beta$. Therefore, if the
values of $\beta$ and $\gamma$ are fixed, the characteristic length
should be also fixed. In our simulations, we give those values as
$\beta=2.0$ 
and $\gamma=2\beta/a_1=2.0/\pi$, so that $k_0=1.0$. 
On the other hand, one of the central subjects in this paper is the
$\alpha$ dependence of domain patterns.
When $\alpha$ becomes
large, the $k$ region with positive $\eta_{\bm{k}}$ broadens
and many modes of $\eta_{\bm{k}}$ grow. Therefore, the surface of
domains can be rough for large $\alpha$.

In this paper, we consider a descending field as the external field
\begin{equation}
  h(t)=h_{\rm ini}-vt,
\label{eq:h}
\end{equation}
where $h_{\rm ini}$ is the initial external field, which should be equal
to or larger than the saturating field. The value of the saturating
field can be estimated from a simulation under ascending field as
follows. After a domain pattern is formed spontaneously under zero
field, an increasing field is applied. At a certain value of $h$, the
domains with negative $\phi(\bm{r})$ disappear. Then, we consider this
value as the saturating field. Although the saturating field depends on
$\alpha$, we set $h_{\rm ini}=1.5$ for all values of $\alpha$ in our
simulations, so that $h_{\rm ini}$ is larger than the saturation fields
for all $\alpha$.

The details of the numerical procedure is as follows. As the initial
condition, $\phi(\bm{r})$ is given by random numbers in the interval
$1.0< \phi(\bm{r}) < 1.1$ at $t=0$. The external field decreases
from $h_{\rm ini}$ to $0$, following Eq.~(\ref{eq:h}). Once the external 
field becomes zero at $t_0=h_{\rm ini}/v$, the field remains zero for
$t_0 \le t \le 2t_0$. Namely, we stop the calculation at $2t_0$. The
time evolution is calculated by using a semi-implicit method. In other
words, while we use the exact solutions for the linear and field terms,
the second order Runge-Kutta method is applied to the calculation for
the nonlinear term. The semi-implicit method enables us to use a rather
large time interval: $\delta t=0.1$ in the simulations below. The
simulations are performed on a $512\times 512$ lattice with periodic
boundary conditions.

\pagebreak

\section{\label{sec:fast} Fast quench}

In this section, we focus on the fast-quench case where the sweep rate
of the field is $v=10^{-2}$. 
The scale of the field sweep rate will be elucidated in Sec.~\ref{sec:cri}.
Considering the results in
Ref.~\onlinecite{kudo}, we expect that some kinds of sea-island
structure should appear in this case. It will be shown how the
characteristics of domain patterns depend on the anisotropy parameter
$\alpha$. The characteristics are demonstrated by the snapshots of
domain patterns, the total number of domains, average domain area, and
domain-area distributions.

\begin{figure}
\includegraphics[width=8cm]{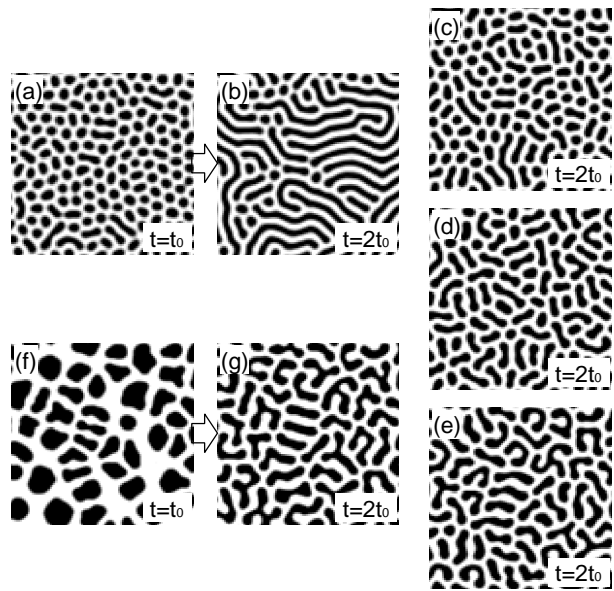}
\caption{\label{fig:snap-2} Domain patterns under zero field after a fast
 quench ($v=10^{-2}$), which display the dependence on the parameter
 $\alpha$: [(a) and (b)] $\alpha=1.5$, (c) $\alpha=2.0$, (d) $\alpha=2.5$,
 (e) $\alpha=3.0$, and [(f) and (g)] $\alpha=3.5$. They are the patterns at
 $t=2t_0$ except for (a) and (f); these two are the ones at $t=t_0$.
 The size of the snapshots
 is $96\times 96$, although the system size is $512\times 512$. The
 value of $\phi(\bm{r})$ is positive and negative in the white and black
 areas, respectively.}
\end{figure}

Figure~\ref{fig:snap-2} shows the domain patterns under zero field. The
white and black areas correspond to positive and negative
$\phi(\bm{r})$, i.e., up and down spins, respectively. 
For $\alpha=1.5$ and $\alpha=3.5$, the domain patterns considerably
change after the external field becomes zero at $t=t_0$. By
contrast, little additional change in domain patterns can be seen for
$\alpha=2.0$, $2.5$, and $3.0$, so that the snapshots for them are
displayed only for $t=2t_0$.
When $\alpha=1.5$, black domains connect with each other due to the exchange
interactions [see Figs.~\ref{fig:snap-2}(a) and \ref{fig:snap-2}(b)]. 
In the other cases (i.e., for $\alpha=2.0$, $2.5$, $3.0$, and $3.5$), the
domains do not tend to connect because of the dipolar interactions. 
However, the exchange interactions prefer connection of domains, and it
is preferable in the viewpoint of energy. When $\alpha$ is small, the
exchange interactions are relatively large and domains can connect with
each other, though the coalescence of domains causes temporary energy
loss of the dipolar interactions.
For $\alpha=3.5$, after large black domains appear
[Fig.~\ref{fig:snap-2}(f)], their shape changes to be like an island with
the same characteristic width as in other cases [compare
Fig.~\ref{fig:snap-2}(g) with
Figs.~\ref{fig:snap-2}(c)--\ref{fig:snap-2}(e)]. 

From Fig.~\ref{fig:snap-2}, one should notice the following:
the larger $\alpha$, the more inhomogeneous width of black
domains. Namely, some domains are partly narrow or thick when $\alpha$
is large. The property can be explained by the linear-growth rate,
Eq.~(\ref{eq:eta}). When $\alpha$ becomes large, the maximum value of
$\eta_{\bm{k}}$ also becomes large. Then, the $k$ region where
$\eta_{\bm{k}}>0$ broadens. Therefore, $\phi_{\bm{k}}$'s with different
values of $\bm{k}$ grow and make the domain patterns with inhomogeneous
thickness.
The large black domains in Fig.~\ref{fig:snap-2}(f) are also explained in
the same way. After some time, Fig.~\ref{fig:snap-2}(f) changes into 
Fig.~\ref{fig:snap-2}(g) because a structure composed of domains whose
characteristic length is $2\pi/k_0$ is more stable than the one composed of
large round domains.

\begin{figure}
\includegraphics[width=8cm]{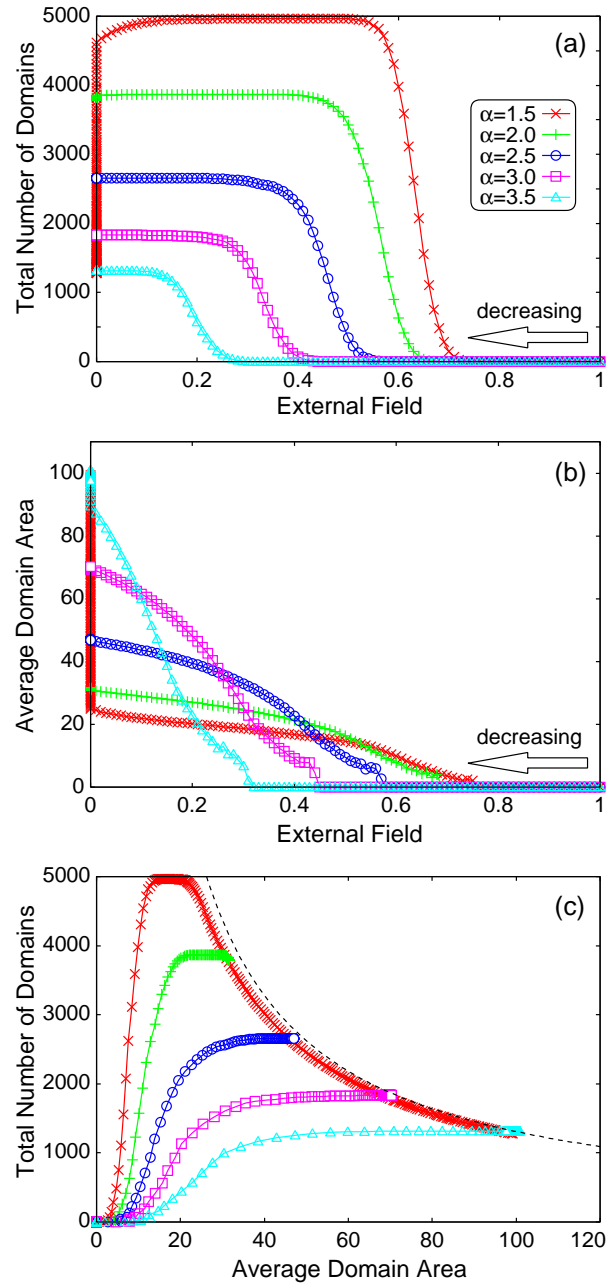}
\caption{\label{fig:stat-2} (Online color) 
The field dependence of (a) the total number
 of domains $N_{\rm tot}$ and (b) the average domain area 
$A_{\rm av}$; (c) the $A_{\rm av}$-$N_{\rm tot}$ graph.
The broken line in (c) expresses the curve of 
$A_{\rm av}N_{\rm tot}=512^2/2$, where the total areas of black and
 white domains are equal. The calculation was performed under a rapidly
 decreasing field ($v=10^{-2}$) until $t=t_0$ and then under zero field
 until $t=2t_0$. The number of domains means the number of
 negative-$\phi(\bm{r})$ (black) domains. The domain area is the number
 of grid points that have negative values of $\phi(\bm{r})$.}
\end{figure}

The field dependence of the total number $N_{\rm tot}$ and average
area $A_{\rm av}$ of negative-$\phi(\bm{r})$ (black) domains is shown
in Figs.~\ref{fig:stat-2}(a) and \ref{fig:stat-2}(b). Under a decreasing
field, $N_{\rm tot}$ starts increasing at a certain field and stops
increasing at another low field. In the plateau region of $N_{\rm tot}$,
each domain grows and $A_{\rm av}$ keeps increasing. When $\alpha=1.5$,
$N_{\rm tot}$ starts to decrease after the plateau region. Moreover, for
$\alpha=1.5$, $N_{\rm tot}$ and $A_{\rm av}$ keep decreasing and
increasing, respectively, after the external field becomes zero. This
behavior is caused by connection of domains as mentioned above:
cf. Figs.~\ref{fig:snap-2}(a) and \ref{fig:snap-2}(b).
If the calculation is continued for $t\ge 2t_0$, the values of 
$N_{\rm tot}$ and $A_{\rm av}$ will decrease and increase more,
respectively. 
On the other hand, when $\alpha=3.5$, $A_{\rm av}$ increases also after
the external field reaches zero, although $N_{\rm tot}$ does not change
under zero field. This means that each domain just grows without
increase of the number of domains. 
For $t\ge 2t_0$, $A_{\rm av}$ does not increase when $\alpha=3.5$.
In fact, $A_{\rm av}$ of $\alpha=3.5$ 
has the maximum value at a certain time between $t_0$ and $2t_0$. 

Figure~\ref{fig:stat-2}(a) also enables us to know that the larger
the $\alpha$, the 
lower the value of the external field where the first black domain
appears. This fact implies that spins cannot easily flop when $\alpha$
is large. In other words, for large $\alpha$, the minima of 
$H_{\rm ani}$ are low and the transition between negative- and 
positive-$\phi(\bm{r})$ states does not easily occur. Another point we
notice 
about Fig.~\ref{fig:stat-2}(a) is that the larger the $\alpha$, the smaller
the $N_{\rm tot}$. This can be explained by the linear growth rate,
Eq.~(\ref{eq:eta}). As mentioned above, the $k$ region where
$\phi_{\bm{k}}>0$ broadens when $\alpha$ becomes large. Therefore, the
size of each domain can become large for large $\alpha$. If the size of
each domain is large, the number of domains should be small.

Figure~\ref{fig:stat-2}(c) shows the $A_{\rm av}$-$N_{\rm tot}$
graph. The broken line is the curve of $A_{\rm av}N_{\rm tot}=512^2/2$,
where the total numbers of the black and white domains are equal. The
curve for each $\alpha$ starts at $(A_{\rm av},N_{\rm tot})=(0,0)$ and
approaches the broken line. The curve for $\alpha=2.0$ and
$\alpha=2.5$ do not reach the broken line, which means that there is a
remanent magnetization.
 
\begin{figure*}
\includegraphics[width=14cm]{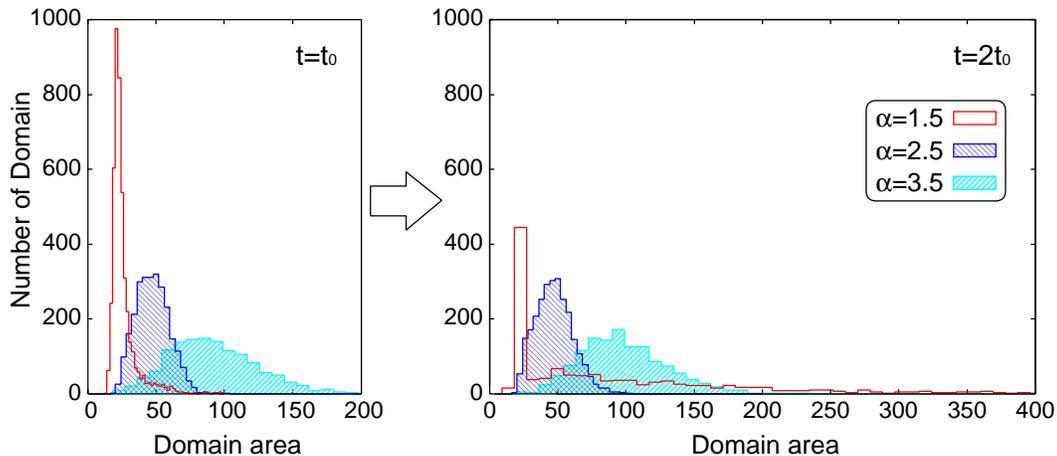}
\caption{\label{fig:ds-2} (Online color) 
Domain-area distributions at $t=t_0$ and $t=2t_0$. 
The left, center, and right histograms at each time are for $\alpha=1.5$,
 $\alpha=2.5$,  and $\alpha=3.5$, respectively.
The domain area is the number
 of grid points that have negative values of $\phi(\bm{r})$.}
\end{figure*}

Figure~\ref{fig:ds-2} displays the domain-area distributions at  $t=t_0$
and $2t_0$ for $\alpha=1.5$, $2.5$, and $3.5$. 
The distribution for $\alpha=1.5$ at $t=t_0$ has a sharp peak at a
certain small value of domain area. The peak falls and a long tail
appears at $t=2t_0$. This change reflects the connection of domains for
$\alpha=1.5$ as mentioned above. For $\alpha=2.5$ or $\alpha=3.5$, the
shape of the each distribution is similar to a Gaussian distribution and
its change between $t=t_0$ and $t=2t_0$ is small. Here, we note that the
larger the $\alpha$, the broader the distribution appears. It can also be
explained by the linear-growth rate.

\section{\label{sec:slow} Slow quench}

In this section, we show our results for the slow-quench case where the
field sweep rate is $v=10^{-4}$. Considering the results in
Ref.~\onlinecite{kudo}, we expect some kinds of labyrinth structure in
this case. The dependence on $\alpha$ of the characteristics of domain
patterns is demonstrated by the snapshots of domain patterns, the total
number of domains, and average domain area. Here, domain-area
distributions are not displayed since each domain can grow too long
under the slowly descending field.

\begin{figure}[b]
\includegraphics[width=8cm]{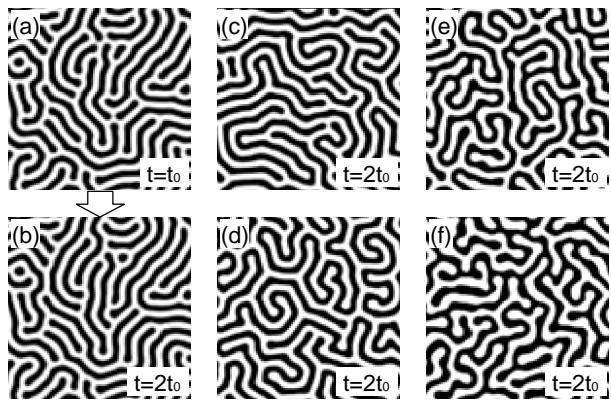}
\caption{\label{fig:snap-4} Domain patterns under zero field after a slow
 quench ($v=10^{-4}$), which display the dependence on the parameter
 $\alpha$: [(a) and (b)] $\alpha=1.5$, (c) $\alpha=2.0$, (d) $\alpha=2.5$,
 (e) $\alpha=3.0$, and (f) $\alpha=3.5$. They are the patterns at
 $t=2t_0$ except for (a) at $t=t_0$.}
\end{figure}

\begin{figure}
\includegraphics[width=8cm]{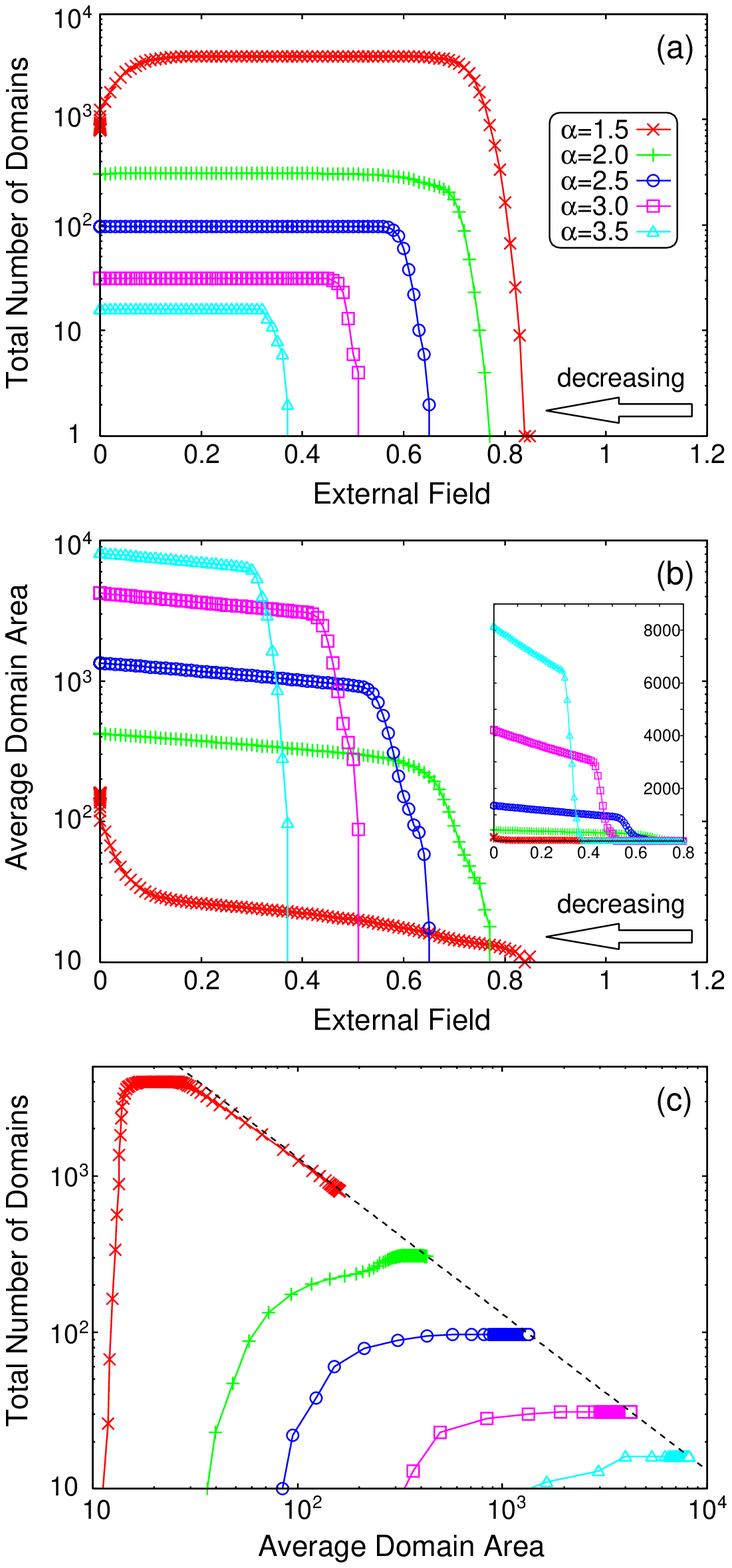}
\caption{\label{fig:stat-4} (Online color) 
The field dependence of (a) the total number
 of domains $N_{\rm tot}$ and (b) the average domain area 
$A_{\rm av}$; (c) the $A_{\rm av}$-$N_{\rm tot}$ graph.
The calculation was performed under a slowly 
 decreasing field ($v=10^{-4}$) until $t=t_0$ and then under zero field
 until $t=2t_0$. 
The data are displayed on a semilogarithmic scale in (a) and (b) 
and on a log-log
 scale in (c), while the inset of (b) shows the same data as its main
 panel on a normal scale. 
The broken line in (c) expresses the curve of 
$A_{\rm av}N_{\rm tot}=512^2/2$. }
\end{figure}

Figure~\ref{fig:snap-4} shows the domain patterns under zero field. In
contrast to the fast-quench case, the domain patterns for $\alpha=1.5$
do not change so much between $t=t_0$ and $t=2t_0$. 
It is because the connection of domains has almost
finished by $t=t_0$.
In fact, the connection of many small domains happens also in this case
for $\alpha=1.5$ (see Fig.~\ref{fig:stat-4}). 
On the other hand, when $\alpha$ is large, the width of black domains is
inhomogeneous. This property is the same as the fast-quench case. We
should note another property: There are many branches of black domains
for large $\alpha$. It is related to the inhomogeneity of domain
width. As black domains grow, they bifurcate at the place where the
domain width becomes large.

The field dependence of the total number $N_{\rm tot}$ and average
area $A_{\rm av}$ of black domains is shown on a semilogarithmic scale in
Figs.~\ref{fig:stat-4}(a) and \ref{fig:stat-4}(b). Also, in this
slow-quench case, $N_{\rm tot}$ decreases after the plateau region for
$\alpha=1.5$ because of the connection of domains. Though $N_{\rm tot}$
decreases and $A_{\rm av}$ increases after the external field becomes
zero for $\alpha=1.5$, they change very little for $t\ge 2t_0$.
Comparing Fig.~\ref{fig:stat-4} with Fig.~\ref{fig:stat-2}, we find two
things. 
First, the difference between both data for $\alpha=1.5$ is rather
small. Second, the larger the $\alpha$, the larger the difference. 
The dependence 
on the field sweep rate was discussed in Ref.~\onlinecite{kudo}
only in the specific case of $\alpha=2.5$. Although the
discussion is basically applicable for other values of
$\alpha$, little difference appears for small $\alpha$ because
spins can flop easily regardless of the field sweep rate.

The inset of Fig.~\ref{fig:stat-4}(b) shows the field dependence of
$A_{\rm av}$ on   normal scale. For $\alpha$ larger than $2.0$, 
$A_{\rm av}$ grows linearly in the low field region. 
The linear-growth region of
$A_{\rm av}$ is within the plateau region of $N_{\rm tot}$. This means that
the magnetization linearly decreases in that region as the external field
decreases. 

Figure~\ref{fig:stat-4}(c) shows the $A_{\rm av}$-$N_{\rm tot}$ graph on
a log-log scale. The end of each curve is almost on the broken line
which expresses the zero magnetization. In other words, the remanent
magnetization is almost zero in this case.

\section{\label{sec:cri} Criterion for the domain structure and standard
 parameter for the field sweep rate}

Now, let us consider the dependence on the field sweep rate. 
Some discussion about it was given in
Ref.~\onlinecite{kudo}. Namely, sea-island and labyrinth structures
appear under zero field for  rapidly and slowly decreasing fields,
respectively. Such behavior was explained by the concept of
crystallization: For a fast-quench case, high ``supersaturation'' lowers
the nucleation energy, while the nucleation energy is high for a 
slow-quench case.\cite{kudo}
When the nucleation energy is small, many domains appear at once and
cannot grow so long. By contrast, when the nucleation energy is large,
a small number of domains appear and can grow very long.
However, these discussions were just qualitative ones, and no standard
parameter responsible to the field sweep rate has been provided. 

In order to predict the domain pattern in experiments, 
a criterion about the dependence on the field sweep rate should
be elucidated. 
Since the proper time scale of each sample is different, the balance
between the proper time scale and the field sweep rate must be
considered to give the criterion. The proper time scale is related to
$L_0$ of Eq.~(\ref{eq:A-C}), while we set $L_0=1$ in the simulations.
First, we should scale out $L_0$ and a parameter $\beta$ in
Eq.~(\ref{eq:A-C}). If we write 
$\tilde{t}\equiv L_0\beta t$, then Eq.~(\ref{eq:A-C}) becomes 
\begin{equation}
 \frac{\partial \phi (\bm{r})}{\partial \tilde{t}}
=  \tilde{\alpha} \lambda(\bm{r}) [\phi(\bm{r})-\phi(\bm{r})^3]
+\nabla^2\phi(\bm{r})
-\tilde{\gamma}\int {\rm d}\bm{r}' \phi(\bm{r}') G(\bm{r},\bm{r}')
+\tilde{h}(\tilde{t}), 
\end{equation}
where $\tilde{\alpha}=\alpha/\beta$, $\tilde{\gamma}=\gamma/\beta$, and
\begin{equation}
 \tilde{h}(\tilde{t})=\frac{h_{\rm ini}}{\beta}-\frac{v}{L_0\beta^2}
 \tilde{t}.
\end{equation}
Here, let us call $\tilde{v}\equiv v/(L_0\beta^2)$ a scaled sweep rate
of the external field.
Next, we have to decide the crossover sweep rate $v_c$ as well as the
scaled one $\tilde{v}_c$. Here, crossover means the crossover from a
sea-island structure to a labyrinth structure. Then, the value of
$v/v_c$ becomes a rough standard in distinguishing sea-island 
and labyrinth structures.
If $L_0$ and $\beta$ of a sample could be measured in experiments, $v_c$
of the sample would be obtained from
$\tilde{v}=v/(L_0\beta^2)$. Then, we can predict the pattern for an
actual $v$ from the value of $v/v_c$.

\begin{figure}
\includegraphics[width=8cm]{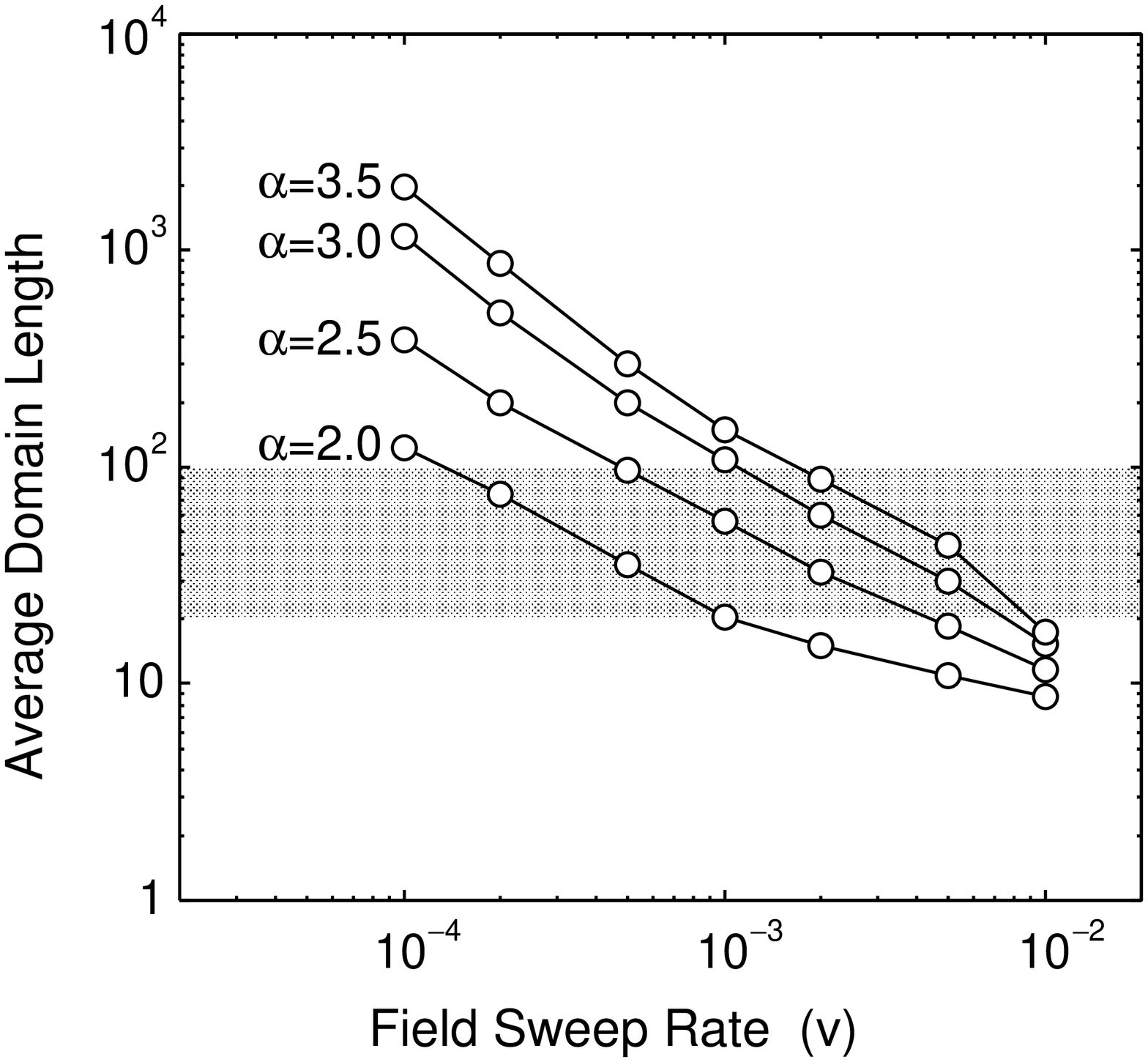}
\caption{\label{fig:v-length}  
Dependence of the average domain length at $t=2t_0$ on the field sweep
 rate $v$. The average domain length is calculated from the average
 domain area and the average domain width (see text).
The domain pattern looks like labyrinth and sea-island structures above
 and below the gray patched area, respectively. 
}
\end{figure}
We need to consider how to distinguish sea-island and labyrinth
structures. Let us introduce the average domain length as a quantity to
distinguish them, which can be obtained by dividing the average domain
area by the average domain 
width. For each $\alpha$ and $v$, the average domain
width was calculated from the autocorrelation function at $t=2t_0$. In
Fig.~\ref{fig:v-length}, the average domain length is plotted against
the field sweep rate for $\alpha=2.0$, 2.5, 3.0, and 3.5. Recalling the
snapshots in Figs.~\ref{fig:snap-2} and \ref{fig:snap-4}, we can expect
labyrinth and sea-island structures above and below the gray patched
area, respectively. Here, we define $v_c=10^{-3}$ in this case, although
it is not a clear crossover point. Then, the crossover sweep rate in
another case is estimated by 
$v_c=\tilde{v}_cL_0\beta^2=2.5\times 10^{-4}\times L_0\beta^2$.

Now, let us introduce a parameter $\kappa=\log_{10}(v/v_c)$. Then, we can
expect that a sea-island structure appears for about $\kappa\ge 1$ and
labyrinth one for about $\kappa\le -1$. 
In fact, the experimental results of Ref.~\onlinecite{kudo} 
support the expectation: Sea-island structure appeared for 
$2\times 10^5$ Oe/s and labyrinth structure for $10$ Oe/s. 
Although we do not know the value of $v_c$ in that case, it is certain 
that the difference between the values of $\kappa$ for sea-island and
labyrinth structures is more than 2 [$=1-(-1)$].  
Therefore, we suggest that the
parameter $\kappa$ should be a useful standard parameter to distinguish
sea-island and labyrinth structures. 

\begin{figure}
\includegraphics[width=8cm]{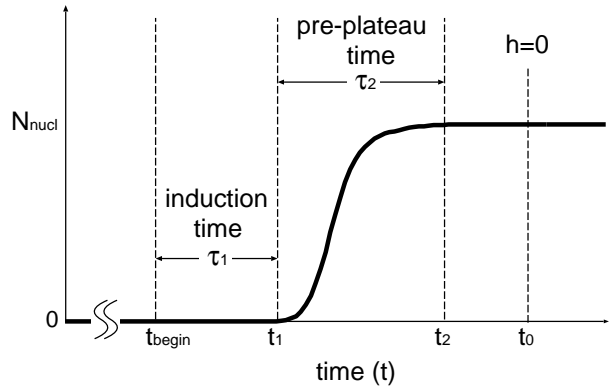}
\caption{\label{fig:t-N}  
Schematic picture about the characteristic times in the nucleation process.
See text for the details.
}
\end{figure}
Here, we explain a linear decay on the logarithmic scale in the 
field sweep-rate dependence of the average domain length 
($\ell_{\rm av}$) in Fig.~\ref{fig:v-length}. 
Let us recall that the data in Fig.~\ref{fig:v-length} are confined to 
the cases where the total number of domains does not change after the 
external field vanishes [see Figs.~\ref{fig:stat-2}(a) and 
\ref{fig:stat-4}(a)].
$\ell_{\rm av}$ is inversely proportional to the number of domains 
($N_{\rm tot}$), which is equal to that of the created nuclei 
($N_{\rm nucl}$) since no connection of domains occurs in the cases 
investigated in 
Fig.~\ref{fig:v-length}. As $N_{\rm nucl}$ is proportional to 
the degree of supersaturation ($\Delta F$), the evaluation of 
$\ell_{\rm av}$ is eventually reduced to that of $\Delta F$.
Now, we consider two characteristic times needed to explain
Fig.~\ref{fig:v-length}: the induction time ($\tau_1$) and the preplateau 
time ($\tau_2$). 
Figure~\ref{fig:t-N} illustrates the important stages of nucleation, 
including those times.
The induction time starts at $t_{\rm begin}$, when the 
supersaturation begins, and ends at $t_1$, when the first nucleus 
appears, i.e., $\tau_1=t_1-t_{\rm begin}$. The preplateau time starts at 
$t_1$ and ends at $t_2$, when $N_{\rm nucl}(=N_{\rm tot})$ begins 
to take the plateau value shown in Figs.~\ref{fig:stat-2}(a), 
\ref{fig:stat-4}(a), and \ref{fig:t-N}, i.e., $\tau_2=t_2-t_1$.
Then, $\Delta F$ is divided in two parts, $\Delta F_1$
and $\Delta F_2$: $\Delta F_1$ and $\Delta F_2$ are the partial degrees 
of supersaturation during the induction time ($\tau_1$) and the preplateau 
time ($\tau_2$), respectively.

We now proceed in estimating $\Delta F_1$ by comparing the energies of 
states (A) and (B) without and with a nucleus, respectively:
(A) all spins are up, i.e., $\phi(\bm{r})=+1$ everywhere;
(B) there is a circular nucleus with radius $R_0$ where spins are down 
among up spins. Namely, $\phi(\bm{r})=-1$ and $+1$ inside and outside 
of the nucleus in (B).
In both (A) and (B), the 
anisotropy energy is the same: $H_{\rm ani}=-\alpha S/4$, where $S$ is 
the area of the system. In (A), $H_J=0$, 
$H_{\rm di}=(2\pi/d)\gamma S$, and $H_{\rm ex}=-hS$.
Therefore, the energy for (A) is 
\begin{equation}
H_A=-\alpha S/4 + (2\pi/d)\gamma S-hS.
\label{eq:HA}
\end{equation}
On the other hand, in (B), the dipolar interactions $H_{\rm di}$
can be estimated from the ratio of combinations of spin pairs. 
Since the ratio of combinations is given by the ratio of areas,
\begin{eqnarray}
H_{\rm di}&=&(2\pi/d)\gamma S\left[ \left(\frac{S-\pi R_0^2}{S}\right)^2
+\left(\frac{\pi R_0^2}{S}\right)^2 -2\frac{S-\pi R_0^2}{S}
\frac{\pi R_0^2}{S}\right]\nonumber \\
&=&(2\pi/d)\gamma S-(2\pi/d)\gamma\cdot 4\pi R_0^2\left(
1-\frac{\pi R_0^2}{S}\right).
\end{eqnarray}
Then we have
\begin{equation}
H_B(t)=-\alpha S/4 +(2\pi/d)\gamma S -(2\pi/d)\gamma
\cdot 4\pi R_0^2\left(1-\frac{\pi R_0^2}{S}\right)-h(t)(S-2\pi R_0^2).
\label{eq:HB}
\end{equation}
Rigorously speaking, there is an additional surface energy in (B) 
due to the exchange interaction ($H_J\sim 2\pi\beta R_0$), which is 
negligible in the present situation with $R_0 \ll S$.
When the difference between a pair of energies $H_A$ and $H_B$,
\begin{equation}
\Delta H_1(t) = H_A-H_B(t) = (2\pi/d)\gamma\cdot 4\pi R_0^2
\left(1-\frac{\pi R_0^2}{S}\right)-2\pi R_0^2(h_{\rm ini}-vt),
\label{eq:DH}
\end{equation}
is positive, it quantifies the degree of supersaturation.
Noting the numerical evidence that the induction time 
depends on $L_0$ and $\alpha$ but not on $v$, we consider $\tau_1$ to 
be independent of $v$. Then, $\Delta F_1$ becomes
\begin{equation}
\Delta F_1=\Delta H_1(t_1) =2\pi R_0^2 \tau_1 v,
\label{eq:F1}
\end{equation}
where we used the equality $\Delta H_1(t_{\rm begin})=0$, i.e., there is no 
supersaturation at the beginning. 

On the other hand, the degree of supersaturation accumulating 
between $t_1$ and $t_2$ is generally written by
\begin{equation}
\Delta F_2 = \int^{t_2}_{t_1} {\rm d}t \left| 
\frac{{\rm d}}{{\rm d}t}\Delta H_2(t)\right|,
\end{equation}
where $\Delta H_2(t)$ stands for the energy difference between the 
present state and its ideal saturated one, which correspond to 
multinuclei versions of $H_A$ and $H_B$ in $\Delta H_1$, at arbitrary 
time during the preplateau time.
We may consider $\Delta H_2(t)$ as a monotonically decreasing function 
since the increase of nuclei reduces the degree of supersaturation
and supersaturation should vanish at $t=t_2$. Noting 
$\Delta H_2(t_1)=\Delta F_1$ and $\Delta H_2(t_2)=0$, we have
\begin{equation}
\Delta F_2 = \int^{t_2}_{t_1} {\rm d}t\left( 
-\frac{{\rm d}}{{\rm d}t}\Delta H_2(t)\right)
=[\Delta H_2(t_1)-\Delta H_2(t_2)]=\Delta F_1.
\label{eq:F2}
\end{equation}

From Eqs.~(\ref{eq:F1}) and (\ref{eq:F2}), $\Delta F=2\Delta F_1$.
Recalling the above mentioned relation 
($\ell_{\rm av}\propto N_{\rm nucl}^{-1} \propto \Delta F^{-1}$), 
we have
\begin{equation}
\log_{10}\ell_{\rm av} \simeq C-\log_{10}v,
\label{eq:logl}
\end{equation}
where $C$ is a constant. Equation~(\ref{eq:logl}) shows the linear decay 
on the logarithmic scale in Fig.~\ref{fig:v-length}. 
While this scenario has still room for improvement, it provides an 
essential explanation of the crossover 
between sea-island and labyrinth structures. 
The improved scenario will appear somewhere in the future.

Recently, a similar behavior to Fig.~\ref{fig:v-length} was reported in
quenched ferromagnetic Bose-Einstein condensations (BECs):\cite{saito}
The number of spin vortices in the quenched BEC depends on the quench
time. The following fact was demonstrated in Ref.~\onlinecite{saito}:
For a slow quench, the spin state has nearly a single-domain structure
and no spin vortices appear, and some spin vortices appear for a fast
quench. In other words, the faster the quench, the smaller the domain
structure. The behavior is really consistent with our results. It will be
an interesting problem to investigate the common mechanism of the
dependence on the quench time.

\section{\label{sec:con} Conclusions}

We have discussed the magnetic domain patterns for several values of
$\alpha$ in the cases of both fast quench and slow one. Except for
$\alpha=1.5$, sea-island and labyrinth structures appear for fast- and
slow-quench cases, respectively. When $\alpha=1.5$, domains connect with 
each
other and form a labyrinth (stripelike) structure. On the other hand,
when $\alpha$ is large, domains tend to have many branches and their
width is inhomogeneous. We have also shown how the characteristics of
the domain patterns, i.e., the number of domains and the average domain
area, change under decreasing field. Moreover, we have introduced the
average domain length as one of the quantities which characterize domain
patterns. The dependence of the average domain length shows that the
change from a labyrinth structure to sea-island one is continuous
against the change in the field sweep rate $v$. Despite this fact,
with the use of the crossover sweep rate $v_c$, we propose the following
criterion: A sea-island structure tends to appear when
$\kappa=\log_{10}(v/v_c)$ is larger than about $1$, and a labyrinth one
does when $\kappa$ is smaller than about $-1$.

\begin{acknowledgments}
The authors would like to thank M.~Mino for information about
 experiments and M.~I.~Tribelsky, M.~Ueda, and Y.~Kawaguchi for useful
 comments and discussion.    
 One of the authors (K.K.) is supported by JSPS. 
\end{acknowledgments}

\end{document}